\documentclass[twocolumn,floatfix,showpacs,tightenlines,superscriptaddress]{revtex4-1}
\usepackage{mathtools}
\usepackage{bm}
\usepackage{dsfont,amsthm,amsbsy}
\usepackage{verbatim}
\usepackage{amssymb}
\usepackage{amsmath}
\usepackage{bbm}
\usepackage{graphicx}
\usepackage{epstopdf}
\usepackage{subfigure}
\usepackage{natbib}
\usepackage{epsfig}
\usepackage{amsfonts}
\usepackage{mathrsfs}
\usepackage{sidecap}
\usepackage{lipsum}
\usepackage[toc,page,title,titletoc,header]{appendix}
\usepackage[colorlinks,linkcolor=blue,citecolor=blue,anchorcolor=blue,urlcolor=blue]{hyperref}
\usepackage{hyperref}
\usepackage{resizegather}
\usepackage{tikz}
\usepackage{float}
\usepackage{mathbbol}
\usepackage[normalem]{ulem}
\usepackage{cancel}
\usepackage{upgreek}

\begin{document}

\title{ Valence-Bond Solid phases in the spin-$1/2$ Kekul{\'e}-Heisenberg model}
	\author{Fatemeh Mirmojarabian}
	\email{fmmojarabian@gmail.com}
	\affiliation{Department of Physics, Institute for Advanced Studies in Basic Sciences (IASBS), Zanjan 45137-66731, Iran}
	\author{Saeed S. Jahromi }
	\email{saeed.jahromi@iasbs.ac.ir }
	\affiliation{Department of Physics, Institute for Advanced Studies in Basic Sciences (IASBS), Zanjan 45137-66731, Iran}
	\author{Jahanfar Abouie}
	\email{jahan@iasbs.ac.ir}
	\affiliation{Department of Physics, Institute for Advanced Studies in Basic Sciences (IASBS), Zanjan 45137-66731, Iran}
	\date{\today}

\begin{abstract}
We map out the ground state phase diagram of the spin $ \frac{1}{2} $ isotropic Kekul{\'e}-Kitaev model on the honeycomb lattice in the presence of the Heisenberg exchange couplings. Our study relies on large-scale tensor network simulations based on a graph-based projected entangled pair state (gPEPS) approach and simple tensor update in the thermodynamic limit. We find that on top of the quantum spin liquid (QSL) and conventional magnetically ordered phases which are typical of the Kitaev-Heisenberg model, the Kekul{\'e}-Heisenberg phase diagram hosts two plaquette valance bond solid (VBS) phases with vanishing magnetic order. While the VBS phases preserve the symmetries of the original Hamiltonian, they differ markedly from the Kitaev spin liquid by having decorated plaquette ordering which is distinguished by a plaquette order parameter.
\end{abstract}

\maketitle
\section{Introduction}
\label{sec:Introduction}
The quest for materials in which quantum fluctuations create quantum paramagnetic phases at temperatures below the interaction energy of the system is an active field of research in condensed matter physics\cite{lacroix2013introduction,Norman2016,Balents2010}.
Quantum spin liquids (QSL) and different instances of valence bond solids (VBS), such as plaquette VBS states and resonating valence bonds are among these phases, to name a few. The emergence of these exotic phases of matter is rooted in strong quantum fluctuations that arise in strongly correlated systems such as frustrated magnets and high-$T_c$ superconductors \cite{Savary2017,Imai2016a,Baskaran1987,Senthil2000,Broholm2020}.

QSLs are of particular importance for their distinguished characteristics such as long-range entanglement, emergent gauge symmetry, fractionalized excitations, and topological order \cite{Savary2017}. On the other hand, in VBSs all spins contribute to the formation of decorated dimers (singlets) which breaks the spatial lattice symmetries. This is in contrast to quantum spin liquids  where no such symmetry is broken. In the ideal case, the two spins forming each singlet are maximally entangled, while disentangled with others, and the ground state of VBSs exhibits no long-range entanglement.

It has already been shown that the spin-$\frac{1}{2}$ antiferromagnetic Heisenberg model on lattices with triangular geometries, such as kagome and star lattices, can potentially host QSLs and various competing VBS phases \cite{Majumdar1969,Fouet2003,Mambrini2006,Zhitomirsky1996,Iqbal2011,Ruegg2010}. Quantum fluctuations induced by geometric frustration trigger a highly degenerate manifold of fluctuating spins that can collectively form exotic quantum orders. 
Spin systems with anisotropic exchange interactions such as Kitaev materials with $4d$/$5d$ elements, e.g. Iridium compounds, are other candidate platforms for realizing QSL and VBS states \cite{Rau2016,Kitagawa2018,Singh,Singh2010,Ye,Modic,Comin,Takayama2015,Sheckelton2012,Smaha2020,Matan2010}. 
The magnetic exchange coupling between the local moments of
$ Ir^{+4}$ ions are included in the low-energy effective Hamiltonian, which is characterized by the Kitaev model \cite{KITAEV} on honeycomb lattice and enhanced by an isotropic Heisenberg interaction \cite{Doubble}. 

The existence of Quantum Spin Liquid (QSL) states in the ground state phase diagram of Kitaev materials has been confirmed by various numerical approaches and experimental techniques  \cite{Corboz2014, Chaloupka2013, Chaloupka2010, Sela2014}. However, experimental investigations of these materials have often revealed the presence of magnetically ordered phases alongside the anticipated QSL behavior. These observations suggest that the Kitaev interaction alone are not sufficient to fully capture the magnetic interactions in these materials. Consequently, the effective Heisenberg term that emerges in subsequent leading orders should also be considered to more precisely describe the magnetic properties of these systems \cite{Jackeli2009, Balents2010, Trebst2017, Lee, Witczak, HwanChun, Reuther}.

\begin{figure}[t]
	\includegraphics[width=\columnwidth]{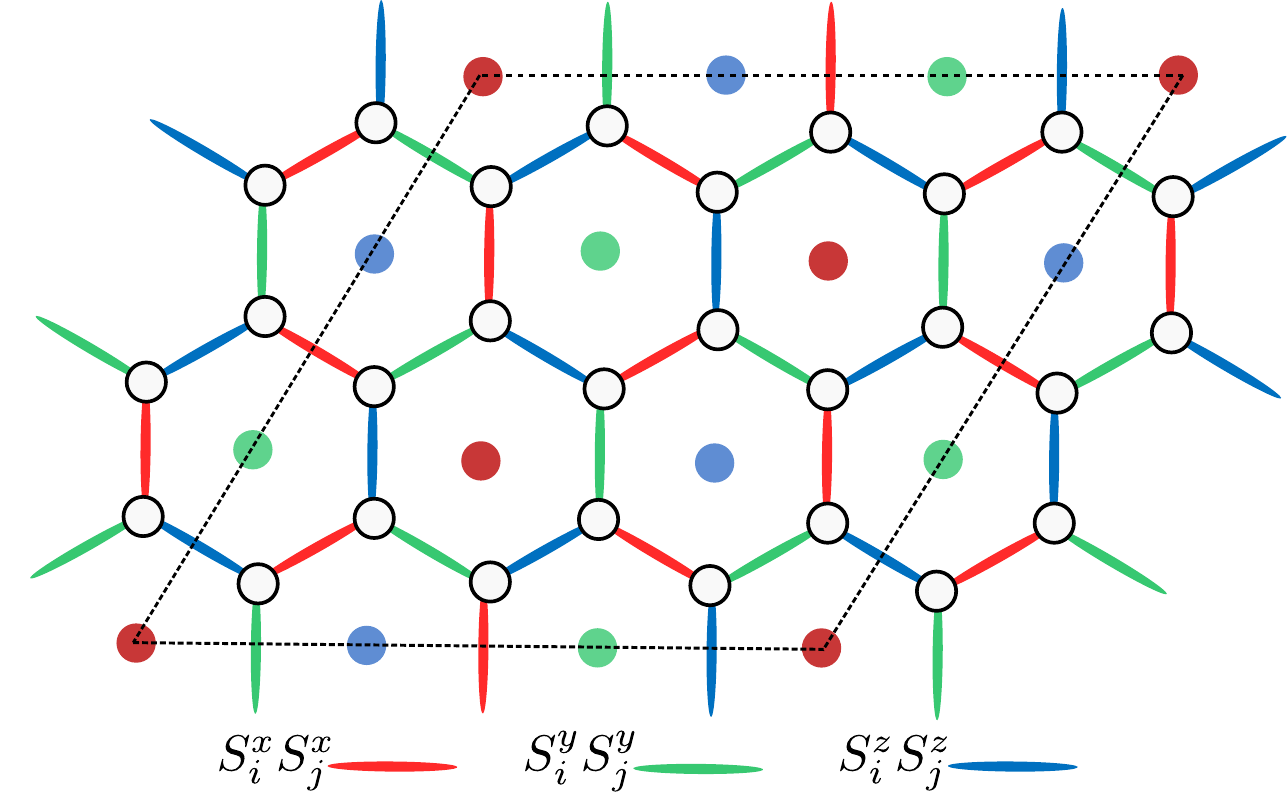}
	\caption{(color online) A schematic illustration of the Kekul\'{e} model on honeycomb lattice. The empty dots sitting on the vertices of the lattice are spin-1/2 particles. The red, green and blue links represent $K^x$, $K^y$ and $K^z$ exchange interactions between two nearest-neighbor spins located at the end of red, green and blue links, respectively. The colored dots at the center of hexagons are used to define the plaquette operators. They show the color of each plaquette (see the subsection \ref{subsec:isoK}). The dotted lines further characterize the translationally invariant 18-site unit cell used in the TN simulations.}
	\label{fig:lattice}
\end{figure}

The Kekul\'{e} model is another variant from the family of Hamiltonians describing Kitaev materials with bond-directional exchange interactions on the honeycomb lattice. The contribution of only two types of Kitaev interaction on each hexagonal plaquette makes the model distinct from the conventional Kitaev-honeycomb lattice (see Fig.~\ref{fig:lattice}).  Although the phase diagram of the  Kitaev-honeycomb model has a gapless region, the ground state of the multi-band Kekul{\'e} model has only one gapless point in the center of the Brillouin zone which is composed of two superimposed Dirac cones \cite{Mirmojarabian2020}.

 Previous studies on the Kekul{\'e} model have shown that ground state phases and phase transitions are distinctly different from those of the original Kitaev model \cite{Moessner,Mirmojarabian2020}. Degenerate perturbation theory in the anisotropic limit of the Kekul{\'e} coupling shows that the low energy effective theory of the system is described by a toric code model on the Kagome lattice. It has also been shown that introducing antiferromagnetic Heisenberg interactions to the model can trigger a quantum phase transition from a QSL state to a magnetically ordered phase through a mechanism known as quantum order by disorder \cite{Moessner}.
 
 By including this term alongside the Kitaev interaction within the Kekul\'{e} Hamiltonian, we investigate the interplay of the Kitaev's bond directional interactions and the Heisenberg's isotropic interactions and map out the full phase diagram of the system.

In this paper we present a comprehensive study of the ground state phases of the Kekul\'{e}-Heisenberg model in the isotropic regime. By incorporating the Heisenberg term alongside the Kitaev interaction within the Kekul\'{e} Hamiltonian, we investigate the interplay between Kitaev bond-directional interactions and Heisenberg's isotropic interactions and map out the full phase diagram of the system. We find that the phase diagram accommodates not only Ferromagnetic (FM) and Antiferromagnetic(AFM) ordered states but also hosts two Valence Bond Solid (VBS) phases with plaquette ordering and two Quantum Spin Liquid (QSL) states.  Our study relies on large-scale tensor network calculations based on an infinite graph-based projected entangled pair state algorithm with simple update. We characterize the underlying phases in the ground state phase diagram by analyzing various physical quantities, such as short-range spin correlations and local magnetization. Additionally, we define appropriate order parameters to distinctly identify different phases.

The paper is outlined as follows: In section \ref{sec:Model} we introduce the Kekul{\'e}-Heisenberg Hamiltonian, and investigate the ground state properties of the system in the pure Kekul{\'e} and Heisenberg limits. We additionally go through our tensor network approach, gPEPS, for simulating the ground state of local Hamiltonian on a 2D honeycomb lattice structure in further detail and study the phase diagram in section~\ref{sec:phasediag}. Finally, Section ~\ref{sec:conclude} is devoted to discussion and outlooks.

\section{Model}
\label{sec:Model}
Let us consider a two-dimensional honeycomb lattice of spin-1/2 particles interacting via the following bond-dependent Kekul{\'e} Hamiltonian: 

\begin{equation}\label{eq:Kekule}
H_{\rm K}=\sum_{\langle i,j\rangle,\alpha} K^\alpha S^{\alpha}_{i}S^{\alpha}_{j}, 
\end{equation}
where $S^{\alpha}$ ($\alpha=x, y, z$) represent the spin-$1/2$ operators and $K$ is the Kekul{\'e} exchange coupling on the nearest neighbor bonds of the lattice. For a better visibility, the bonds with $x, y, z$ Kitaev interactions have been colored respectively with red, green, and blue. In contrast to the standard Kitaev Honeycomb model, only two colors contribute to the construction of each plaquettes in the Kekul\'{e} model. We can further label each plaquette by the color of the outgoing links. This will distinguish three colored triangular sub-lattices obtained by connecting the center of the plaquettes with the same color (See Fig.~\ref{fig:lattice}).

The Kekul\'{e}-Heisenberg (KH) model is further described as an interplay of the Kekul\'{e}-Kitaev and Heisenberg interactions, which naturally arise in the real materials. In the presence of the Heisenberg interaction, the Hamiltonian of the model is given by:
\begin{equation}\label{eq:Kekule-Heisenberg}
H_{\rm KH}=H_{\rm K} + J\sum_{\langle i,j\rangle} \mathbf{S}_{i}\cdot\mathbf{S}_{j},
\end{equation}
where $J$ is the Heisenberg exchange coupling between all nearest-neighbor spins. 

The interplay of the Kekul{\'e} and the Heisenberg interactions can lead to a rich phase diagram for the underlying Hamiltonian \eqref{eq:Kekule-Heisenberg}. Before delving into the ground-state phase diagram of the KH model, we first provide a brief overview of the ground state properties of the Hamiltonian \eqref{eq:Kekule-Heisenberg} in the extreme limits of pure Heisenberg model, and isotropic and anisotropic Kekul{\'e} models.

\subsection{Heisenberg limit ($J \neq0 $, and $K^ \alpha=0 $)}\label{subsec:Heisenberg}

In the regime where the Kekul{\'e} term in the Hamiltonian \eqref{eq:Kekule-Heisenberg} is switched off, i.e. when $K=0$, the model reduces to the Heisenberg model on the Honeycomb lattice. For $J<0$, the model represents the ferromagnetic (FM) interaction and the ground state of the system possesses an FM long-range ordering with a finite magnetization and a positive spin-spin correlation. In this case, the excitations of the system are magnons with the low-energy dispersion $\varepsilon(k)\sim k^2$. In contrast, for the $J>0$ couplings, the ground state has a N{\'e}el ordering with a non-zero staggered magnetization. Due to the staggered arrangement of the spins in the N{\'e}el state, all short-range correlations have a negative sign. In this case the excitations are magnons with linear low-energy dispersion $\varepsilon(k)\sim k$.

\subsection{Isotropic Kekul\'{e} limit ($J=0$, and $K^\alpha=K\neq0$)}
\label{subsec:isoK}

In the limit of vanishing Heisenberg exchange coupling, i.e. when $J=0$, the KH Hamiltonian reduces to the Kekul{\'e} model \eqref{eq:Kekule}. Similar to the Kitaev honeycomb model, the Kekul{\'e} model can be exactly solved by utilizing the following spin-Majorana fermion transformation: $S_j^{\alpha}=\frac{\mathrm{i}}{2}b_j^{\alpha}c_j$, where ${\rm i}$ stands for imaginary, and $b_j^\alpha$ with $\alpha = x, y, z$ and $c_j$ are Majorana fermion operators that create/annihilate a Majorana fermion of type $b^\alpha$ and $c$ at cite $j$, respectively \cite{Moessner,Mirmojarabian2020}.

Following the straightforward algebra proposed by Kitaev\cite{KITAEV}, the Majorana representation of the Hamiltonian \eqref{eq:Kekule} reads
\begin{equation}\label{eq:Hmajorana}
H_{\rm {KM}}=K\frac{\mathrm{i}}{4}\sum_{\langle i, j\rangle} \frac{u_{i,j}^{\alpha}}{2} c_{i}c_{j},
\end{equation} 
where $u_{i,j}^{\alpha}=\mathrm{i}b_{i}^{\alpha}b_{j}^{\alpha}$ is the bond operator corresponding to the connection link between the two nearest-neighbor sites $i$ and $j$. These operators possess the following properties: I) different bond operators commute with each other, i.e., $[u_{i,j}^{\alpha},u_{i,j}^{\beta}]=\delta_{\alpha\beta}$, II) they commute with the Hamiltonian $H_{\rm{KM}}$, and III) their square is equal to identity, implying that the eigenvalues of the bond operator $ u_{ij} $ can only be $+1$ or $-1$. This constraint gives rise to a fixed $\mathbb{Z}_{2}$ gauge property on every bond of the honeycomb lattice, indicating that each bond can exist in one of the two states, i.e., the eigenvalue of the bond operator is either $ u_{ij}=+1 $ or $ u_{ij}=-1 $. To avoid any confusion and facilitate the visualization of bond operators, each bond is assigned a specific direction. A convention would be that for even $i$s and odd $j$s, the eigenvalue of the bond operator is $+1$.
 
The Kekul{\'e} model \eqref{eq:Kekule} has also three integrals of motion defined as $W^{\gamma}=\prod_{(i,j)\in \gamma} u_{i,j}$ around the plaquettes of the honeycomb lattice. Here, $\gamma$ is the color of the underlying plaquette which is distinguished by the color of the outgoing link. The plaquette operators square to identity, i.e, $(W^\gamma)^2=1$ and therefore have $w^\gamma=\pm1$ eigenvalues. Besides, they commute with each other, i.e. $[W^{\gamma}, W^{\gamma'}]=\delta_{\gamma\gamma'}$, and with the Hamiltonian \eqref{eq:Kekule}, i.e., $[H, W^\gamma]=0$. The ground state of the Kekul{\'e} model \eqref{eq:Hmajorana}, according to the Lieb theorem \cite{Lieb}, is in the zero-flux sector
corresponding to the configuration with $w^{\gamma}=1$ for all plaquettes.

The Fourier transform of the Hamiltonian \eqref{eq:Hmajorana} is given by\cite{Mirmojarabian2020}:
\begin{equation}\label{eq:HpureMajorana}
H(\textbf{q})=\frac{\mathrm{i}}{2}\sum_{q}\varPsi_{\textbf{q}}^{T} A(\textbf{q}) \varPsi_{-\textbf{q}},
\end{equation}
where $A(\textbf{q})$ is a skew-symmetric $6\times 6$ matrix presented in the Appendix~\ref{skew}, and ${\varPsi_{\textbf{q}}^{T}=(c_{1\textbf{q}},c_{2\textbf{q}},c_{3\textbf{q}},c_{4\textbf{q}}
,c_{5\textbf{q}},c_{6\textbf{q}})}$ is the transpose of the vector $\varPsi_{\textbf{q}}$.
As seen, in the Majorana fermion language, the interacting Kekul{\'e} spin system is reduced to a system of non-interacting Majorana fermions with the exactly solvable Hamiltonian \eqref{eq:HpureMajorana}.  
By diagonalizing the matrix $A({\textbf q})$, the energy spectrum of the Hamiltonian \eqref{eq:HpureMajorana} is readily obtained. 
The spectrum is gapless and the dispersion is composed of two superimposed Dirac cones at the center of the Brillouin zone \cite{Mirmojarabian2020,Moessner}. It is worth mentioning that in the Kitaev model the Dirac cones appear at $K$ and $K^{'}$ points \cite{KITAEV}, however, in the Kekul\'{e} model the crossing of the Majorana bands occurs at the $\Gamma$ point. This has a significant impact on the stability of the Majorana modes, i.e., while the Kitaev model exhibits a gapless region, the Kekul\'{e} model has solely one gapless point at the isotropic point with $K^\alpha=K$.

In the presence of both the Kekul{\'e} and Heisenberg interactions concurrently, it is no longer possible to explore the ground-state properties of the system (Eq.~\eqref{eq:Kekule-Heisenberg}) analytically, and therefore we should employ numerical techniques. 
To do this, we employ the infinite projected entangled-pair state (iPEPS), and  gPEPS techniques which are effective tensor network (TN) approaches to examine the ground-state features of two-dimensional quantum lattices. In the next section, we briefly sketch how the tensor networks may be utilized to generate the ground state phase diagram of the Kekul{\'e}-Heisenberg model on the two-dimensional lattices.

\section{Tensor Network Method}
\label{sec:method}
Tensor network methods are among the state-of-the-art numerical techniques which can potentially be used for providing efficient representation for the ground state of quantum many-body systems (QMBS). In recent years, TN algorithms have been used extensively for the study of many exotic phases of matter ranging from spin liquids to frustrated magnets and even fermionic QMBSs.\cite{ORUS2014,ORUS2019,biamonte2017tensor,Verstraete_2008}. The fundamental concept behind TN techniques is to efficiently write the wave function of local many-body Hamiltonians based on their entanglement structure \cite{ORUS2014}. Matrix product state (MPS) and projected entangled-pair state (PEPS) are two well-known examples of TN states for representing the ground state wave function of the 1d and 2d QMBS, respectively. The infinite version of these TN states, i.e., the iMPS and iPEPS \cite{ORUS2014,Verstraete_2008,Verstraete2006} have also been used extensively for the study of various QMBS in the thermodynamic limit. 

\begin{figure}[t]
	\includegraphics[width=\columnwidth]{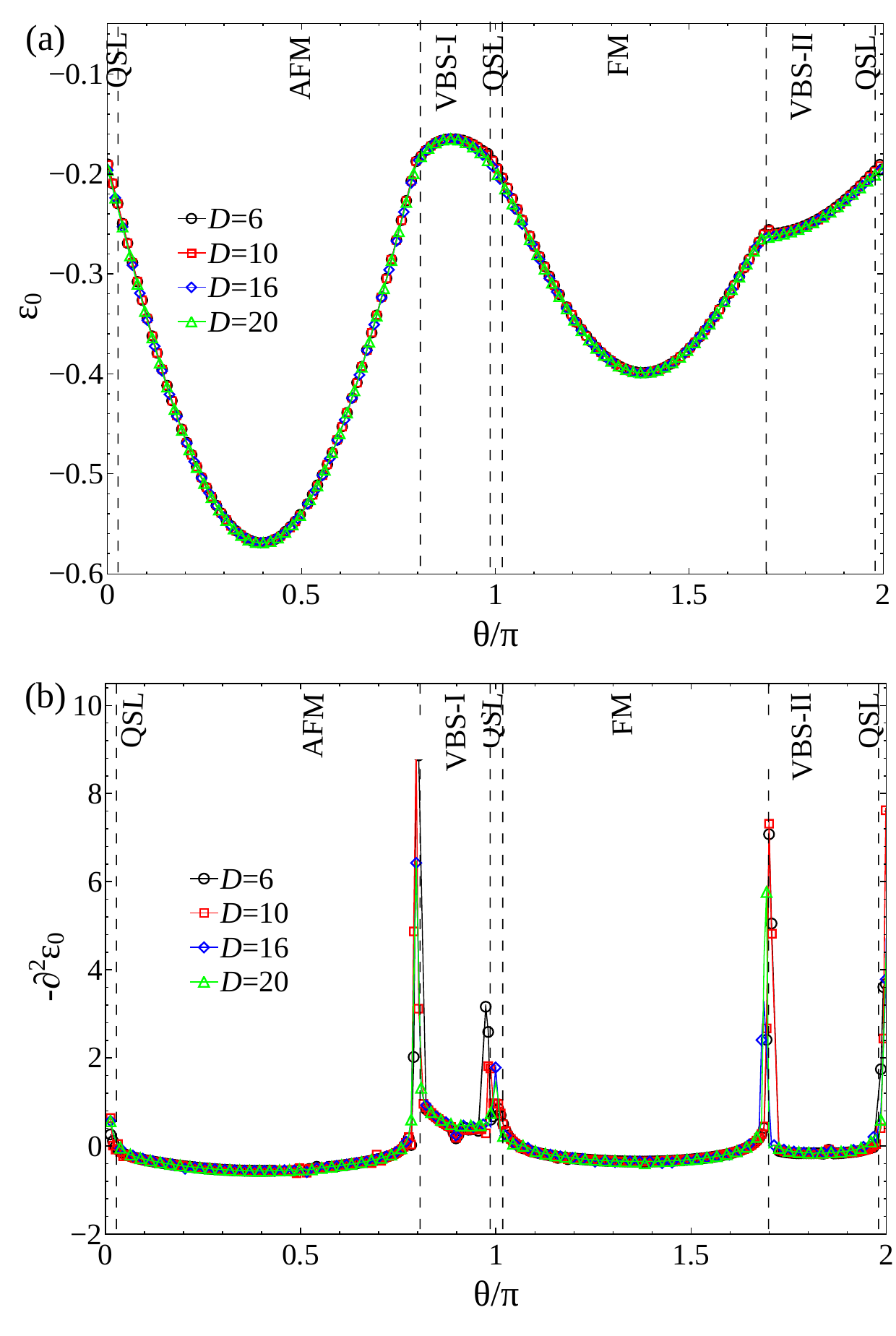}
	\caption{(color online) (a): The ground state energy per site in the Kekul\'{e}-Heisenberg model for multiple bond dimensions $D$. (b): The second derivative of the ground state energy with respect to $\theta$. 
	}\label{fig:energy}
\end{figure}

In general, the TN representation for the ground state of a local lattice Hamiltonian can be approximated efficiently by variational methods \cite{Corboz2016} or through approaches based on evolving the system in imaginary-time \cite{Jordan2008,Orus2009,Phien_2015}. Despite their considerable potential and effectiveness, these strategies are constrained by implementation difficulties related to dimensionality and lattice geometry. More specifically, the contraction of the infinite tensor networks can be highly challenging and cumbersome depending on the lattice structure. One, therefore, resorts to approximation techniques such as boundary MPS techniques or approaches based on corner transfer-matrix renormalization group for the contraction of the infinite many-body wave function and calculation of expectation values and correlations.

\begin{figure}[t]
	\includegraphics[width=1.0\columnwidth]{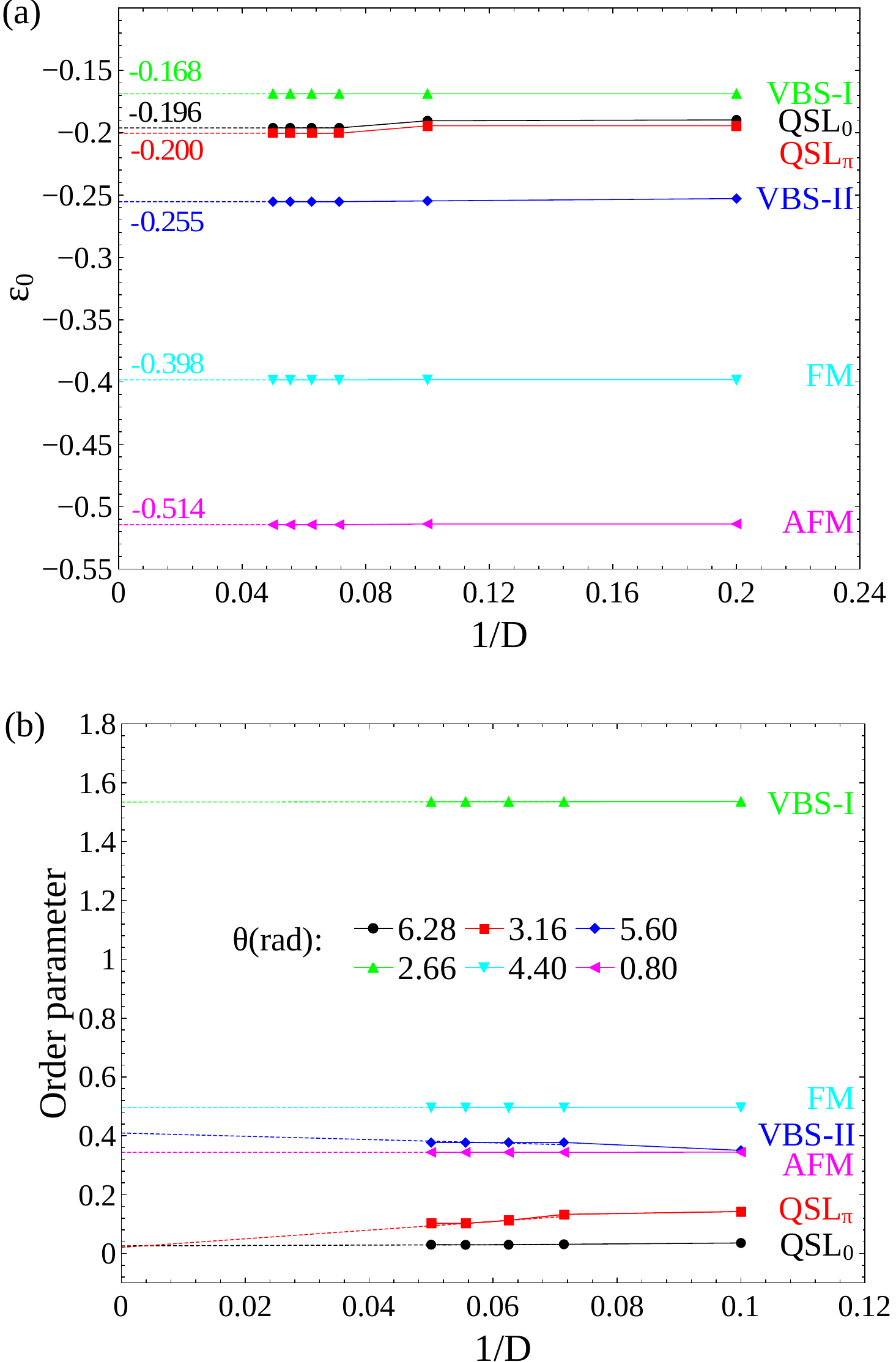}
	\caption{(color online) (a) Scaling of the ground state energy per-site as a function of the inverse bond dimension ($1/D$) for representative points within distinct phases across the phase diagram. (b) Scaling of the order parameters for the same points in (a). For the spin liquid phases, scaling of the total magnetization is presented as a signature of vanishing local ordering for the QSL states. For VBS phases the scaling lines show the VBS order parameter as defined in Eq.\ref{Eq:VBS-order parameter}. For FM and AFM phases the lines illustrate total and staggered magnetization, respectively. The convergence of both energy and order parameters across various $\theta$ values indicates the stability of the phases in the thermodynamic limit.}
	\label{fig:energy_scaling}
\end{figure}

\begin{figure*}[t]
	\includegraphics[width=\textwidth]{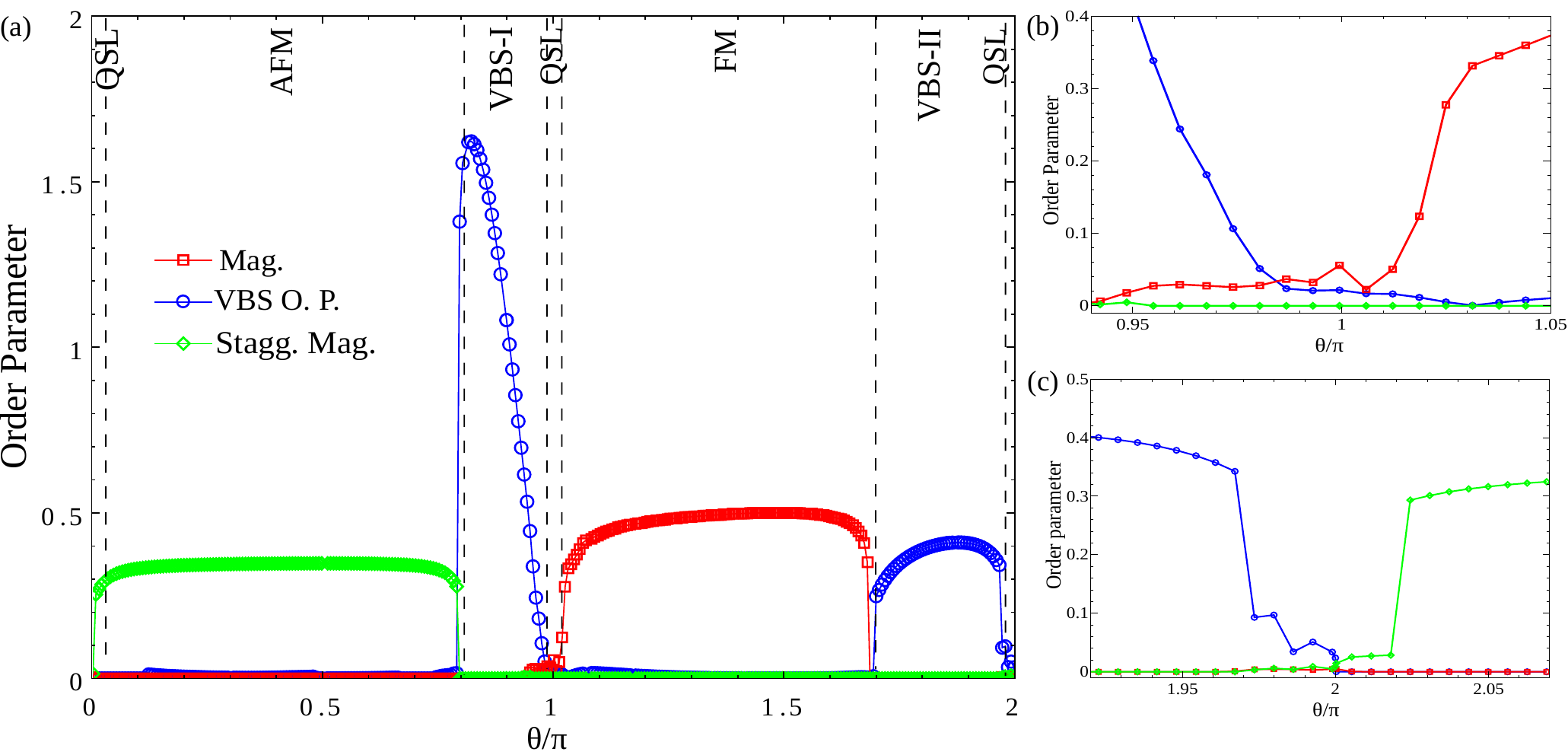}
	\caption{(color online) (a)Different order parameters versus $\theta$ computed by means of gPEPS with $D=20$. The Magnetization (red), staggered magnetization (green) and VBS order parameter (blue) are nonzero in the FM, AFM N{\'e}el and VBS phases, respectively. (b) and (c) Zoomed-in order parameters around $\theta=\pi$ and $\theta=2\pi$, respectively.
	}\label{fig:O.P}
\end{figure*}

Another alternative approach to tackle the simulation of QMBSs is to consider the lattice structure as a directed graph and use the graph-based infinite PEPS algorithm, i.e. gPEPS technique \cite{Jahromi2019}, to simulate the ground state of local Hamiltonians on complicated lattice structures. The gPEPS method codifies the connectivity information of the underlying lattice into the so-called structure-matrix (SM) and overcomes the geometrical implementation difficulties \cite{Jahromi2019,Jahromi2020}. The tensor updates can then be fully automatized by going through the columns of the SM in a methodical way that doesn't depend on the geometry. Using the simple-update (SU) approach based on imaginary-time evolution, the ground state of local nearest-neighbor Hamiltonians can be approximated efficiently for any lattice and dimensionality. Within the SU scheme, the environments of local tensors are supplied by bond matrices produced via local singular-value decomposition (SVD), which offer an approximation of the environment and correlations surrounding local tensors at the mean-field level. While more precise approaches such as full-update (FU) \cite{Corboz2012,Corboz2013,Corboz2014} have been developed to capture the whole system correlation, their implementation, and truncation are generally constrained by lattice shape. Yet, it has been demonstrated that the mean-field approximation of the environment in the SU algorithm is fairly accurate for higher-dimensional systems and thermal states, making the SU a quite accurate solution in these cases \cite{Jahromi2019, Jahromi2020}. This is mainly because by increasing the dimension and lattice connectivity (lattice coordination number), the correlations in the system are distributed among more degrees of freedom, resulting in less entanglement per bond in practice (due to entanglement monogamy).

In the followings, we use the gPEPS technique to simulate the ground state of the Kekul{\'e}-Heisenberg model on the infinite honeycomb lattice and map out the full phase diagram of the system in the thermodynamic limit. We performed the TN simulations on translationally invariant unit-cells with different sizes. The model requires a minimum unit cell containing $18$ sites to encompass all three independent plaquettes, shown in Fig.~\ref{fig:lattice}. However, we performed simulations by considering larger cells to make sure we capture the correct physics in the thermodynamic limit.

The maximum virtual bond dimension achievable within our available computational resources is ${D_{\text{max}}= 20}$. Additionally, we employed simple update based on imaginary time evolution and second order Suzuki-Trotter decomposition, initiating with $\delta\tau=10^{-1}$ and progressively reducing to $10^{-3}$ while allowing a maximum of $3000$ iterations for each $\delta\tau$. 
Furthermore, we monitored both the energy and the singular values obtained during the simple update to ensure algorithm convergence.  We terminated the process once a threshold of $10^{-16}$ was reached for the convergence of the singular values. The tensors were initialized randomly, and the ground state for each coupling was calculated several times from different random initial states. This approach ensures that the algorithm consistently converges to a truly stable state in the thermodynamic limit.

In the next section, we will present the gPEPS phase diagram and discuss the underlying phases and phase transitions.

\section{Ground state phase diagram}
\label{sec:phasediag}

In order to scan the full parameter space and extract the phase diagram of the KH model, we set ${K=\cos\theta}$ and ${J=\sin\theta}$, and computed the ground state of the KH model for ${\theta \in [0, 2\pi]}$ with the gPEPS technique. To capture the phase boundaries and distinguish different phases, we have plotted the ground state energy, $\varepsilon_0$, and its second derivative with respect to $\theta$ ($-\partial^2\varepsilon_0/\partial\theta^2$) in Fig.~\ref{fig:energy}.  
The phase boundaries are best detected in the plot of the second derivative of energy that reveals the existence of six distinct regions separated by vertical dashed lines in Fig.~\ref{fig:energy}-(b). Fig.~\ref{fig:energy} further contains the energy results for several bond dimensions $D$. Deep inside each phase, the energy and its second derivative do not depend on $D$ (also see the scaling of the ground state energy as a function of $1/D$ plotted in Fig.~\ref{fig:energy_scaling}-(a)), but near the transition points, the values are slightly different, and higher bond dimensions tend to produce more precise results.
\begin{figure*}[t]
	\includegraphics[width=\textwidth]{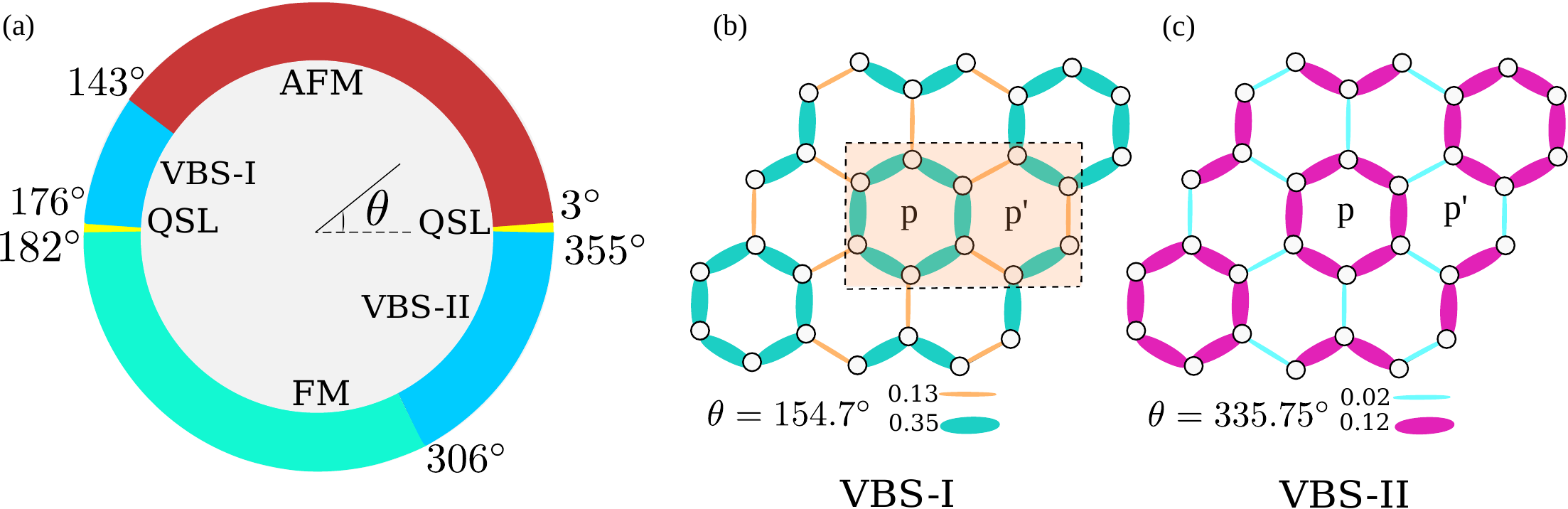}
	\caption{(color online) (a): The ground state phase diagram of the spin $\frac{1}{2}$ Kekul{\'e}-Heisenberg model on honeycomb lattice. Here we  defined $J=\sin\theta$ and $K=\cos\theta$. On the top (bottom) of the circle the Heisenberg interaction is dominant and the system is in the AFM (FM) phase. On the left and right of the circle, the Kekul{\'e} interaction is dominant and the system is in the QSL phase. In the rest of the circle the interplay between Kitaev and Heisenberg interactions leads to two plaquette VBS phases. (b) and (c), Nearest-neighbor spin correlation patterns in the VBS-I and VBS-II phases, respectively. Thickness of the bonds indicates the strength of the correlations. The plaquettes with strong correlations on all bonds are labeled with $p$. Furthermore, $p'$ represents the plaquettes with alternating strong-weak correlation pattern. The shaded area in panel (b) represents an example of two neighboring $p$ and $p'$ plaquettes that are used in Eq.~\eqref{eq:VBS-order-parameter} to calculate the Plaquette VBS Order parameter (PO).} 
	\label{fig:phasediagram}
\end{figure*}
In order to pinpoint the nature of each phase distinguished by the energy and its derivative, we first investigate the ground state properties of the system at the extreme limits of $\theta=0$ and $\pi$, where the KH Hamiltonian \eqref{eq:Kekule-Heisenberg} reduces to the Kekul\'{e} model with QSL ground state. Analysis of the energy and its derivative shows that the QSL phases span for $\theta \in [355^{\circ}, 3^{\circ}]$ and $\theta \in [176^{\circ}, 182^{\circ}]$ regions around the pure antiferromagnetic and ferromagnetic Kekul{\'e} points, respectively. These narrow QSL regions are distinguished with vertical dashed lines in the energy plot in Fig.~\ref{fig:energy}. In Fig.~\ref{fig:energy_scaling}-(b) we have illustrated our TN results on the magnetization of QSL states, $\sum_{i}   \lvert\langle \textbf{S}_{i} \rangle\rvert $, as a function of $1/D$ (see the red and black curves). The behavior of the magnetization in QSL phases shows that, it approaches zero by increasing the bond dimension $D$ which is in agreement with the QSL nature of the Kekul{\'e} ground state at $\theta=0,~\pi$ points. 
 
The KH Hamiltonian \eqref{eq:Kekule-Heisenberg} further corresponds to the antiferromagnetic (ferromagnetic) Heisenberg model at $\theta=\pi/2$ ($\theta=3\pi/2$) with non-zero magnetic orders. As previously pointed out in Sec.~\ref{subsec:Heisenberg} the ground state of the antiferromagnetic Heisenberg model on the honeycomb lattice is a N{\'e}el state with anti-parallel alignment of the spins. This AFM ordering can be best detected with staggered magnetization, i.e., $\sum_{i} (-1)^{i} \langle \textbf{S}_{i} \rangle $. Fig.~\ref{fig:O.P} (the green curve) demonstrates a nonvanishing staggered magnetization for $\theta \in [3^{\circ}, 143^{\circ}]$ which also includes the $\theta=\pi/2$ point, and distinguishes a large AFM region in the phase diagram of the KH model. Alternatively, a large FM region can be detected for $\theta \in [182^{\circ}, 306^{\circ}]$ (see Fig.~\ref{fig:O.P}, the red curve) which spans around the pure ferromagnetic Heisenberg model at $\theta=3\pi/2$ (Also see the phase diagram, illustrated in Fig.~\ref{fig:phasediagram}-(a)).  

So far, we have identified four phases out of the six total phases in the phase diagram of the KH model which consists of two QSL regions and two magnetically ordered phases with FM and AFM ordering (Fig.~\ref{fig:phasediagram}-(a)). From Fig.~\ref{fig:O.P}, we find that the two remaining regions with $\theta \in [143^{\circ}, 176^{\circ}]$ and $\theta \in [306^{\circ}, 355^{\circ}]$, both have vanishing total and staggered magnetization. While this behavior is reminiscent of the lack of long-range magnetic orders in the QSL states, a careful analysis of the short-range correlations, $\langle \textbf{S}_{i}  \cdot\textbf{S}_{j}\rangle$, on the different bonds of the honeycomb lattice confirms the existence of a valance-bond-solid state with plaquette ordering that is composed of a strong-week pattern of bond correlations. A snapshot of these VBS phases has been illustrated in Fig.~\ref{fig:phasediagram}-(b,c). This is in sharp contrast to the Kekul{\'e} spin liquid states which all links of the honeycomb lattice have uniform correlation. 

 In order to distinguish the VBS state in the phase diagram, we define the following VBS order parameter:
\begin{equation}
PO=|\sum_{\langle i,j\rangle  \in p} \langle \textbf{S}_{i}  \cdot\textbf{S}_{j}\rangle -\sum_{\langle i,j\rangle  \in p'} \langle \textbf{S}_{i}  \cdot\textbf{S}_{j}\rangle |,
\label{eq:VBS-order-parameter}
\end{equation}
where the summations run over all the nearest neighbor links of the two nearest neighbor hexagonal plaquettes $p$ and $p'$ as shown in the shaded region of the Fig.~\ref{fig:phasediagram}-(b). Here $p$ refers to the plaquette with strong correlations on all bonds and $p'$ represents the plaquette with alternating strong-weak correlation pattern, as depicted in Fig.~\ref{fig:phasediagram} (b) and (c). The plaquette VBS order parameter (blue curve in Fig.~\ref{fig:O.P}) is zero in all regions except for the VBS phases. Let us emphasize that the bond correlations on the links of the $p$ and $p'$ plaquettes differ in magnitude. As a result, the two sums in Eq.~\eqref{eq:VBS-order-parameter} do not cancel each other out, and the VBS order parameter will not be zero in the VBS phases of the Kekul{\'e}-Heisenberg Hamiltonian.
 
 The VBS phase appears in two separate regions in the phase diagram: one is between the QSL and AFM phases, which we call VBS-I, and the other between the QSL and the FM, which we call VBS-II. In both phases, the plaquette order parameter is nonzero, and the lattice symmetry is broken. However, since they appear in two regions with different couplings, their bond correlation and consequently the plaquette order parameter are different in magnitude. This can be best seen from the blue curve in Fig. \ref{fig:O.P}. The strength of the spin-spin correlation on the links of the lattice for two arbitrary angles ($\theta=154.7^{\circ}, \theta=335.75^{\circ}$) deep inside the VBS-I and VBS-II regions has been reported in Fig.~\ref{fig:phasediagram}-(b,c) indicating their difference at the local level. To characterize the differences between the two VBS states, we have further calculated the ground state fidelity for two wave functions, each corresponding to one of the VBS phases. The ground state fidelity is defined as       
\begin{equation}
F(\psi_1 ,\psi_2)=\left | \dfrac{\langle \psi_1  |\psi_2 \rangle}{ \sqrt{\langle \psi_1 | \psi_1 \rangle \langle \psi_2 | \psi_2 \rangle} } \right | ,
\label{Eq:fidelity}
\end{equation}
 where $\psi_1(\psi_2)$ are two un-normalized ground state wave functions in VBS-I and VBS-II phases, respectively. To be more specific, we used the local tensors around the central plaquette $p$ in Fig.~\ref{fig:phasediagram}-(b), (c) for $\theta_1=0.89\pi $  and $\theta_2=1.84\pi$, deep inside the VBS-I and VBS-II phases, respectively and calculated the fidelity per plaquette by contracting the tensors according to TN diagram of the inset of Fig.~\ref{Eq:fidelity}. Here the green triangles at the edges represent the bond matrices of simple update that characterize the surrounding environment of the local tensors at the mean-field level. The tensor network itself corresponds to the overlap between the translationally invariant components of the wave functions in VBS-I and VBS-II. Fig.~\ref{Eq:fidelity} further demonstrates the scaling of fidelity per plaquette versus inverse bond dimension $1/D$ for the overlap of the two VBS states. The vanishing overlap in the thermodynamic limit suggests that the VBS-I and VBS-II states have a different local nature. It is notable that in this calculation, the virtual bond dimension (D) remains identical for both ground state wave functions. However, an alternative method to compute fidelity with varying D for two wave functions is elaborated in Appendix~\ref{App:fidelity}. This approach is suitable to calculate the overlap of the numericaly obtained PEPS wave functions and the well known exact states such as dimer states or other plaquette wave functions. Such overlap calculations can help to gain deeper insight into the local nature of the VBS states which we leave for future studies.

Moreover, the scaling of the ground state energy per site and the order parameters with respect to the inverse of the bond dimension, as depicted in Fig.~\ref{fig:energy_scaling}, demonstrates very good convergence across all regions in the phase diagram. This suggests that the observed phases are stable states in the thermodynamic limit.

\begin{figure}[t]
	\includegraphics[width=\columnwidth]{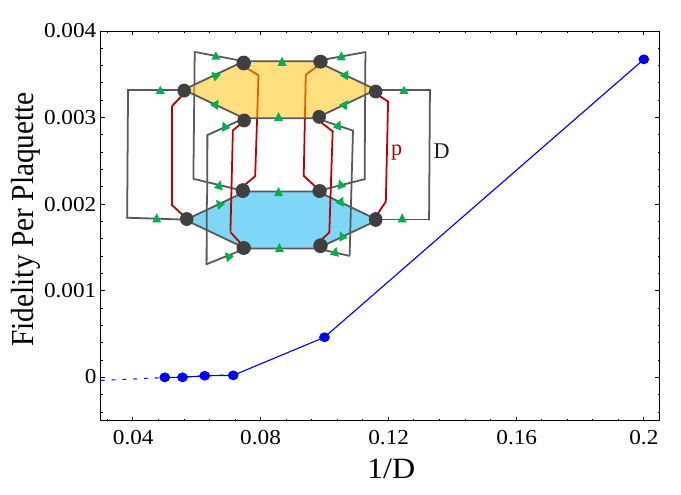}
	\caption{(color online) Scaling of the ground state fidelity per plaquette (Eq~\eqref{Eq:fidelity}) as a function of inverse bond dimension $(1/D)$ for $\theta_1=0.89\pi$  and $\theta_2=1.84\pi$, deep inside the VBS-I and VBS-II phases, respectively. The inset shows local contraction of tensors for calculating the ground state fidelity per plaquette, i.e., the overlap of two wave functions $\psi_1$ and $\psi_2$ in two distinct regions of the phase diagram.}
	\label{fig:fidelity}
\end{figure}

\begin{figure}[t]
	\includegraphics[width=\columnwidth]{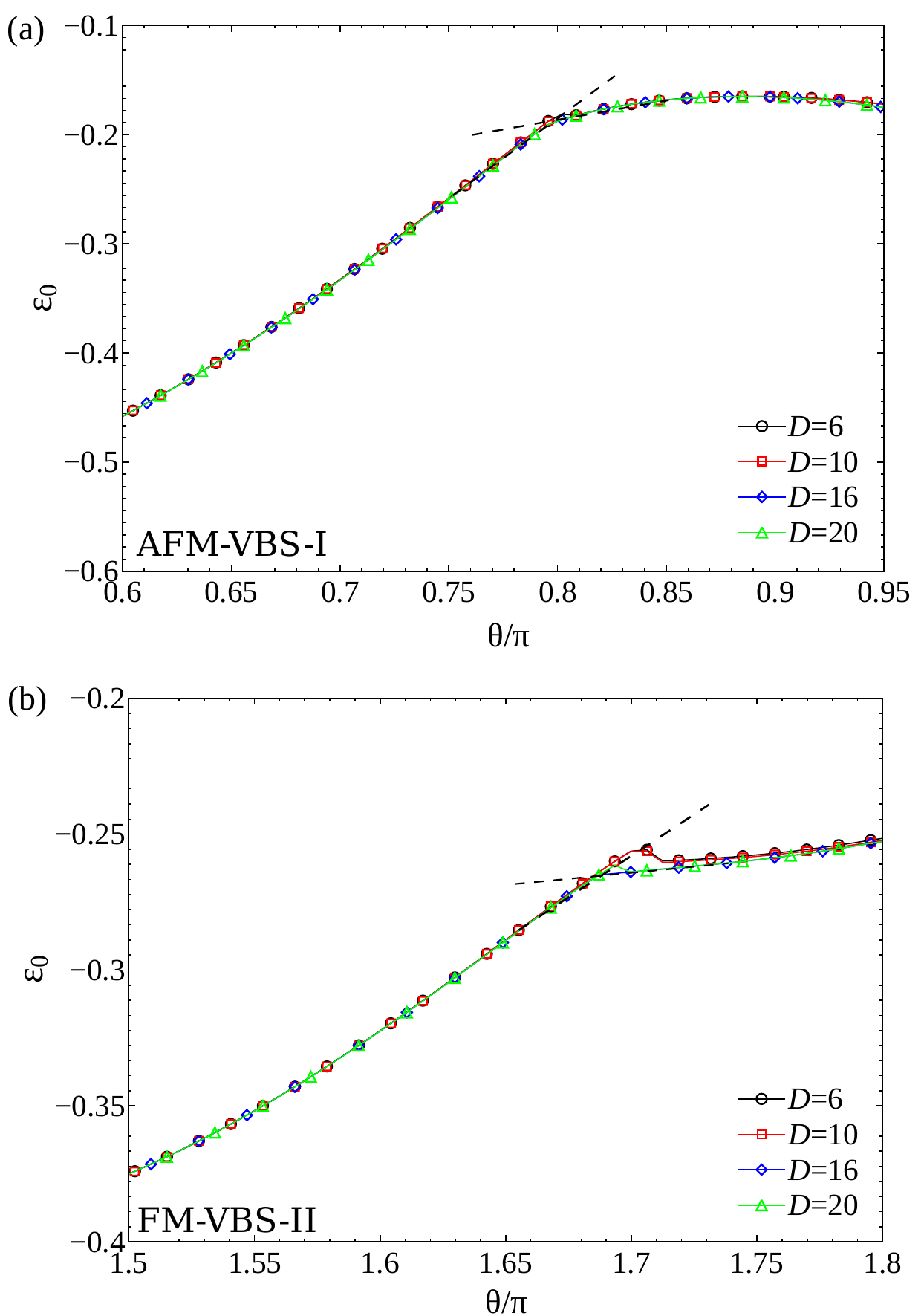}
	\caption{(color online) The zoomed-in plot of the ground state energy per-site for (a) the AFM-VBS-I transition and (b) the FM-VBS-II transitions. The level crossing observed at the critical point (indicated by the dashed lines) signifies a discontinuous jump in the first derivative of the energy which is a characteristic feature of first-order phase transitions.}
	\label{fig:energy-zoom}
\end{figure}

By analyzing the ground state energy of the KH model and its derivative near the phase boundaries, and also investigating the behavior of the order parameters at the transition points, we are able to characterize the nature of quantum phase transitions between different regions of the phase diagram. To characterize the nature of phase transitions more convincingly, we show zoome d-in energy plots for the AFM-VBS-I and FM-VBS-II transitions in Fig.~\ref{fig:energy-zoom} (a) and (b), respectively. The AFM-VBS-I and FM-VBS-II transitions are clearly first order. This can be best seen from the level crossing of the energy fits from both sides of the transition points (see dotted lines in Fig.~\ref{fig:energy-zoom}-(a), (b)), implying a discontinuous first order derivative of the energy which is a signature of first order phase transitions. Detecting the nature of other phase boundaries, particularly the transitions into and out of the Quantum Spin Liquid (QSL) regions, presents challenges. The limited accuracy of our tensor network (TN) simulations at critical points, especially for the gapless QSL states that appear in very narrow regions, restricts our ability to provide a concrete characterization. However, our analysis of the gPEPS ground state energy suggests that the VBS-I-QSL, VBS-II-QSL and QSL to FM and AFM transitions are likely continuous. 
	
Let us wrap up this section by pointing out that an intriguing distinction between the phase diagram of the Kekul{\'e}-Heisenberg and the Kitaev-Heisenberg models on the honeycomb lattice is that while in the former, VBS states, VBS-I and VBS-II, emerge between the QSL and a magnetically ordered phase, AFM or FM, the phase diagram of the latter hosts a zigzag and a stripy magnetic ordering in the vicinity of the QSL regions. This makes the Kekul{\'e}-Heisenberg model a platforms to observe various phases of matter ranging from conventional magnetic states to more exotic phases such as VBSs and QSLs.  

\section{Discussion and outlook}
\label{sec:conclude}

Kitaev materials, such as various Iridate compounds, serve as promising platforms for realizing diverse and intriguing phases of matter, ranging from magnetically ordered states to Quantum Spin Liquids (QSLs). Previous research has established that the interplay between the anisotropic, bond-dependent Kitaev interactions and the naturally occurring Heisenberg exchange coupling in these materials leads to a complex phase diagram that includes QSL phases alongside various magnetically ordered states. In this paper, we demonstrate that a variant of the Kitaev honeycomb models, which features a distinct decoration of bond-anisotropy known as the Kekul{\'e} model, can host a unique and relatively rare family of quantum paramagnetic phases, the plaquette VBS states, stemming from the competition between the Kekul{\'e} and Heisenberg interactions.

Using large-scale tensor network simulations based on infinite projected entangled-pair states optimized using the simple update approach, we mapped out the full phase diagram of the Kekul{\'e}-Heisenberg model on the honeycomb lattice. Our findings reveal that, in addition to the Quantum Spin Liquid (QSL) and magnetically ordered phases, the KH model hosts a plaquette Valence Bond Solid (VBS) state in two distinct regions of the ground state phase diagram. These occurrences of the plaquette VBS state have not been previously observed in standard Kitaev models on different lattice geometries, highlighting the unique impact of the Kekul{\'e} modifications on the system's quantum phase behavior.

Within the accuracy of our analysis, we distinguished a first-order nature for the quantum phase transitions between the FM, AFM and the VBS states. However, due to the limitations of the simple update near critical points, our ability to definitively classify the nature of the phase transitions between the QSL phases and the antiferromagnetic, ferromagnetic, and the Valence Bond Solid (VBS) states was limited. 

Lastly, it's important to emphasize that the phase diagram of the Kekul{\'e}-Heisenberg model differs significantly from that of the standard Kitaev-Heisenberg model. While the Kitaev-Heisenberg model features both Ferromagnetic and Antiferromagnetic ordered states, as well as two additional magnetic states, zigzag and stripy phases, the Kekul{\'e}-Heisenberg phase diagram replaces these zigzag and stripy phases with plaquette VBS phases. This distinction highlights the unique effects of the Kekul{\'e} modifications on the system’s quantum phase behavior, introducing new complexities and phenomena to explore.

\section{Acknowledgments}
This research was supported by the Institute for Advanced Studies in Basic Sciences (IASBS). We also thank A. Langari and M. Kargarian for fruitful discussions.

\appendix
\section{Skew-symmetric  matrix\label{skew}}
In this study, we provide the mathematical representation of skew antisymmetric matrices A(\textbf{q}) that are observed in the Hamiltonian equation \eqref{eq:HpureMajorana}. Let us consider the honeycomb lattice, assuming that its primitive unit vectors are denoted as $a_1 = (1, 0)$ and $a_2 = (1/2, \sqrt{3}/2)$. The A(\textbf{q}) is defined as
\cite{Mirmojarabian2020}:

\begin{widetext}
\begin{equation}\label{eq:AHpureMajorana}
A(\textbf{q})=
\left(\begin{array}{cccccc}      
 0 & -2K& 0 &-2K e^{\mathrm{i}q_{x}} & 0 & -2K\\
2K&0&2K&0&2K e^{\mathrm{i}(q_{x}-\sqrt{3}q_{y})/2}&0\\
0 &-2K&0&-2K& 0&-2Ke^{-\mathrm{i}(q_{x}+\sqrt{3}q_{y})/2}\\
2K e^{-\mathrm{i}q_{x}} &0&2K&0&2K&0\\
0&-2K e^{-\mathrm{i}(q_{x}-\sqrt{3}q_{y})/2}&0&-2K&0&-2K\\
2K&0&2Ke^{\mathrm{i}(q_{x}+\sqrt{3}q_{y})/2}&0&2K&0
\end{array}\right),
\end{equation}
\end{widetext}
where $q_x$ and $q_y$ are the two components of the $\mathbf q$ vector.	The eigenvalues of this matrix are the energy spectra of the system. 
The skew-symmetry exhibited by the matrix $ A(\textbf{q}) $ can be understood as a reflection of the particle-hole symmetry inherent in the Majorana Hamiltonian, which arises from the Majorana condition $ c ^\dagger = c $. As a result of this particle-hole symmetry, all eigenvalues of $ A $ occur in pairs. 
The unperturbed ground state of the  Kekul{\'e} model is a topological state belonging to the BDI symmetry universality class\cite{Moessner,Mirmojarabian2020}.

\section{Calculation of fidelity\label{App:fidelity}}
\begin{figure}[t]
 	\includegraphics[width=\columnwidth]{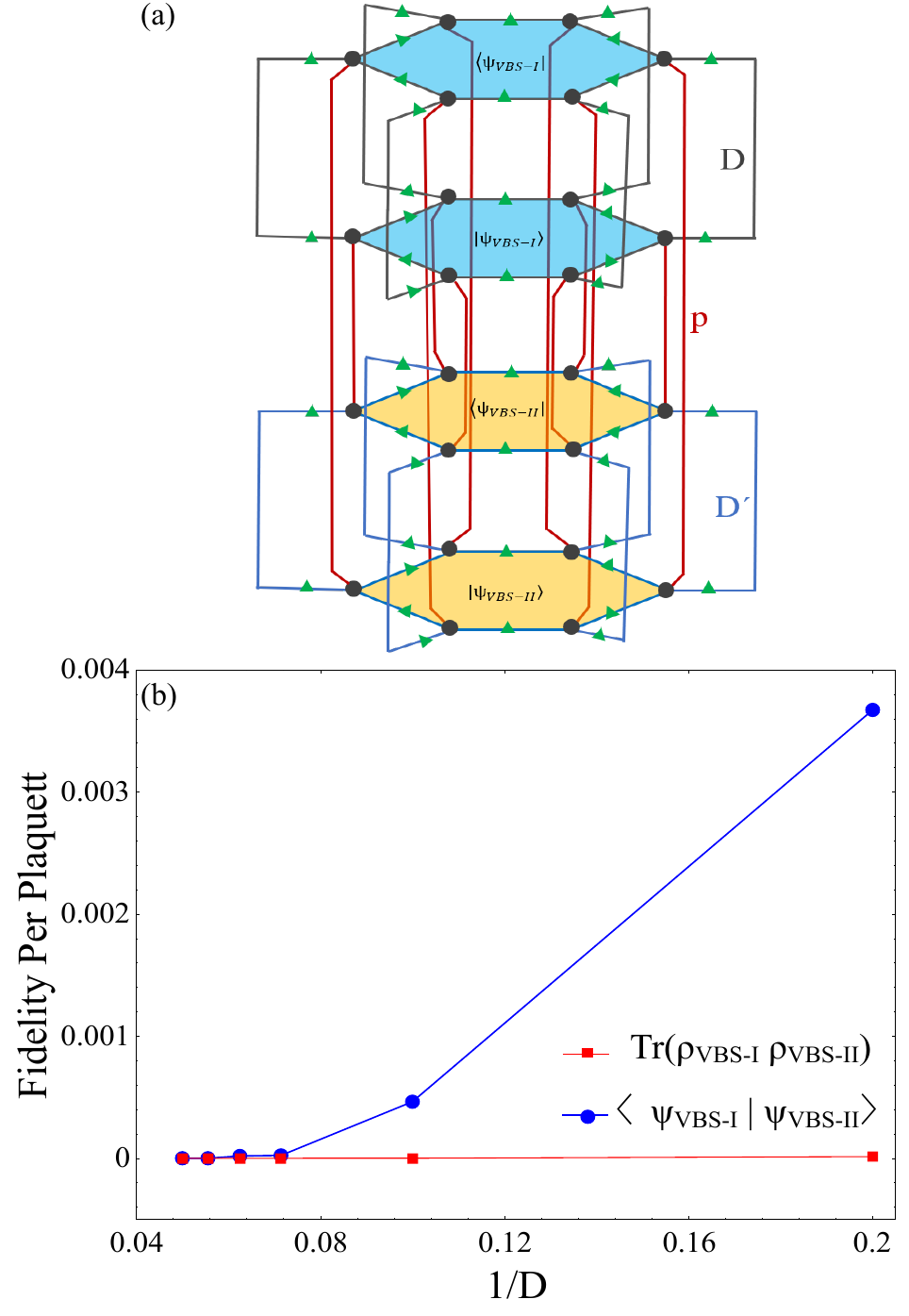}
 	\caption{(color online) (a) Tensor network diagram of the density matrix formalism for calculating the overlap between two wave functions with different virtual bond dimension. (b) Scaling of the ground state fidelity per plaquette from Eq.~\eqref{Eq:fidelity} and the density matrix formalism as a function of inverse virtual bond dimension $(1/D)$ for the same $theta$s as those in Fig.~\ref{fig:fidelity} deep inside the VBS-I and VBS-II phases.} 
 	\label{fig:Trace}
 \end{figure}

As previously discussed in section \ref{sec:phasediag}, in order to calculate the fidelity per plaquette one can proceed with Eq.~\eqref{Eq:fidelity} and the tensor network proposed in the inset of Fig.~\ref{fig:fidelity}. This is basically the overlap between two ground state wave functions of the VBS-I and VBS-II with the same virtual bond dimension. The alternative approach  to study the overlap is to use the density matrix formalism to and calculate $Tr(\rho_{VBS-I} \rho_{VBS-II})$. In this approach we contract the tensor network diagram of Fig.~\ref{fig:Trace}-(a). Similar to the TN diagram in Fig.~\ref{Eq:fidelity} the green triangles on the edges show bond matrices of simple update characterizing environment of local tensors. The advantage of  construction depicted in Fig.~\ref{fig:Trace} for investigating the overlap is that we can contract PEPS tensors with different virtual bond dimension $D$  and $D'$. Let us point out that the overlap from the density matrix formalism is actually equivalent to $|F(\psi_1 ,\psi_2)|^2$ where $F$ is already defined in Eq.~\eqref{Eq:fidelity}.

 Fig.~\ref{fig:Trace}-(b) demonstrates the overlap with density matrix formalism as compared to the result of Eq.~\eqref{Eq:fidelity} as a function of inverse of bond dimension$1/D$. The fidelity overlap from both approaches vanish in the thermodynamic limit suggesting distinct local nature of VBS-I and VBS-II phases.

\bibliography{refs}

\end{document}